# Data and Spectrum Trading Policies in a Trusted Cognitive Dynamic Network

B. Lorenzo, *Member, IEEE*, A. Shams Shafigh, *Student Member, IEEE*, J. Liu, *Student Member, IEEE*, J. González-Castaño, *Senior Member, IEEE*, Y. Fang, *Fellow, IEEE*

*Abstract*—Future wireless networks will progressively displace service provisioning towards the edge to accommodate increasing growth in traffic. This paradigm shift calls for smart policies to efficiently share network resources and ensure service delivery. In this paper, we consider a cognitive dynamic network architecture (CDNA) where primary users (PUs) are rewarded for sharing their connectivities and acting as access points for secondary users (SUs). CDNA creates opportunities for capacity increase by network-wide harvesting of unused data plans and spectrum from different operators. Different policies for data and spectrum trading are presented based on centralized, hybrid and distributed schemes involving primary operator (PO), secondary operator (SO) and their respective end users. In these schemes, PO and SO progressively delegate trading to their end users and adopt more flexible cooperation agreements to reduce computational time and track available resources dynamically. A novel matching-with-pricing algorithm is presented to enable self-organized SU-PU associations, channel allocation and pricing for data and spectrum with low computational complexity. Since connectivity is provided by the actual users, the success of the underlying collaborative market relies on the trustworthiness of the connections. A behavioral-based access control mechanism is developed to incentivize/penalize honest/dishonest behavior and create a trusted collaborative network. Numerical results show that the computational time of the hybrid scheme is one order of magnitude faster than the benchmark centralized scheme and that the matching algorithm reconfigures the network up to three orders of magnitude faster than in the centralized scheme.

## I. Introduction

The rapid growth of advanced wireless devices and services is exacerbating the problem of spectrum scarcity and posing potential challenges for mobile operators, especially in terms of quality of service (QoS) provisioning [1]. Existing solutions for coping with traffic demand focus on investing in additional fixed infrastructure, which is costly from an environment and network perspective [2]. Besides, these solutions rely on conventional cellular infrastructure design built to satisfy peak rates and ignore the dynamic traffic fluctuations that render a significant part of this infrastructure unutilized in space and time. Despite densification efforts to increase spectrum reusability, the licensed spectrum continues to be scarce and its efficient usage will soon approach the theoretical limits [2].

Recently, a new generation of dynamic network architectures (DNAs) has emerged in which users share their connectivities and act as access points for other users in their vicinity, augmenting network capacity [3]-[5]. The high density of users' terminals provides many opportunities for connectivity and network traffic offloading without additional infrastructure cost. Integrating cognitive capabilities into the DNA for spectrum harvesting will facilitate access to additional unused spectrum, both temporally and spatially, to meet growing spectrum demand [3]. In addition, the diversity of data plans[1] provides users with different service capabilities as potential access points. If users with high capabilities outsource these to others (by acting as access points) at critical network operational times, overall network performance can be improved and harvested spectrum can be more efficiently utilized via an increase in frequency reuse. Furthermore, network operators could intelligently share their residual spectrum resource and service capabilities to ensure service delivery and thus, increase their revenue and create new business opportunities. These opportunities result from trading harvested data and spectrum between users and operators.

In this paper, we explore business opportunities in data and spectrum harvesting created by a cognitive dynamic network architecture (CDNA) where primary users (PUs) share their connectivities with secondary users (SUs) for some reward. In CDNA, each SU connects through its preferred PU using the harvested spectrum. The selected PU shares its unused data and acts as an access point for SU transmissions in return for a reward. CDNA creates a new collaborative market for data and spectrum trading and opportunities for revenue sharing among the parties involved (primary operator [PO], secondary operator [SO] and their respective end users). A framework for data and spectrum trading optimization is developed to maximize the utility of each party and satisfy the QoS for SUs. Three approaches are considered: centralized, hybrid and distributed. Each incurs different levels of coordination and revenue sharing. In the centralized approach, the SO performs data and spectrum trading with the PO to satisfy the demands of SUs. The PO then rewards PUs willing to serve as access points for SU traffic. In the hybrid scheme, the SO and PO trade the spectrum but delegate data trading to PUs and SUs. PUs benefit directly from this trading as an incentive to share their resources. Finally, in the distributed scheme, the SO and PO negotiate a revenue share for their cooperation and let PUs and SUs trade the data and spectrum.

To fully exploit the potentials of this architecture and make it highly adaptive to traffic dynamics, we aim to develop distributed mechanisms for data and spectrum trading with low computational complexity. The traffic dynamics result from the activity of the SUs and PUs, creating resource-sharing opportunities. In [4], Shafigh et al. present a genetic algorithm for topology reconfiguration in DNA without cognitive capabilities and show that the optimum topology can track network dynamics and, thus, satisfy users' QoS requirements. However, genetic algorithms are centralized in nature and in our CDNA, both the PO and SO participate in resource allocation. Game theory is a powerful tool for performing distributed resource allocation [6], [7]. In [8] the spectrum-sharing problem between a set of device to device (D2D) pairs and multiple co-located cellular networks is formulated as a Bayesian non-transferable utility overlapping coalition formation game. Some works have applied matching theory to cognitive networks to solve the channel allocation problem [9], [10]. Matching theory seems an attractive framework for resource allocation in cognitive networks where two sets of agents (PUs and SUs) can be matched according to their preferences [35]. Nevertheless, existing works in this area have some limitations. In [9], a one-to-one stable matching game is considered where the utilities of the SUs and PUs are chosen to be identical because the SUs cannot obtain the performance measures of the PUs. In [10] the channel assignment problem is formulated as a many-to-one matching game under the limitation that each primary channel can only be assigned to one SU. Our work overcomes these limitations since CDNA facilitates operator cooperation, and we incorporate pricing into the matching decision to facilitate self-organized data and spectrum trading among multiple SUs and PUs.

---
[1] The data plan limits the amount of data transferred over the network for the duration of the plan. We will use data plan trading and data trading interchangeably through the paper.

Since connectivity opportunities are offered by users in CDNA, to ensure widespread adoption, it is crucial to develop trust mechanisms that encourage trustworthy connections. In cognitive networks, trust mechanisms have been proposed for collaborative spectrum sensing under report falsifying attacks [11], [12]. A user-selection method based on reinforcement learning is presented in [11] to select reliable SUs. Qin et al. [12] proposed a trust-based model and developed a weighted sensing aggregation scheme to remove attackers from the decision-making process. However, a holistic design for a trust management system is missing due to a lack of network architecture.

Our major contributions are summarized as follows:

- A framework for data and spectrum trading optimization which involves a PO, a SO and their respective end users. We develop three schemes—centralized, hybrid and distributed—with new policies for revenue sharing, and derive the optimal behavior of the parties involved. These schemes present different performance and complexity tradeoffs to suit heterogeneous traffic loads and revenue expectations.

- A matching theory based algorithm with pricing to solve the SU-PU association for data and spectrum trading, spectrum allocation and pricing. The algorithm is based on a many-to-one matching game where multiple SUs are assigned to a PU that satisfies their QoS requirements. Pricing is incorporated into the matching as the price of resources is a decisive criterion for SU-PU association. Positive and negative externalities are considered due to changes in demand and supply, which have an impact on data and spectrum availability and, thus, on price.

- A two-stage deterrence-based trust mechanism which encompasses partially distributed (via local physical interactions) and partially centralized (via operators' involvement) trust management. Since perception of trust varies among users, we model behavioral aspects and their impact on trust, and define a behavioral-based access control scheme that encourages consistent behavior through punishment (of misbehaved users) and reward (of well-behaved users). By exploring the properties of trust, we also propose a fully distributed trust mechanism for autonomous evaluation of trust by users.

The rest of this paper is organized as follows. The related work is reviewed in Section II. The network model is discussed in Section III. The data and spectrum trading framework is given in Section IV. New matching algorithms are developed in Section V. The trust relationship model is presented in Section VI. Performance evaluation is given in Section VII. Finally, Section VIII concludes the paper.

## II. RELATED WORK

Novel architectures for cognitive networks based on D2D [8], [13], small cells [14], [15] and multi-hop communications [16], [17] have been proposed to further increase spectrum efficiency. D2D spectrum sharing is intended to offload traffic from the cellular infrastructure when source and destinations are close to each other [13]. Cognitive capabilities can be utilized to mitigate intercell interference between a macrocell and the small cells to deploy heterogeneous spectrum-efficient networks [15]. However, the introduction of a fixed infrastructure significantly increases overall energy consumption and infrastructure cost. Enabling multi-hop communication in cognitive networks can further increase coverage and spectrum efficiency by exploiting locally available channels and support dynamic traffic distributions without additional infrastructure costs [16]. The importance of backup channels and relaying incentives to increase link reliability and robustness in multi-hop cognitive networks is addressed in [18].

Due to the enormous economic value of spectrum, spectrum trading has attracted a lot of interest lately on the aforementioned architectures [18]-[23]. In [19], repeated auctions in the uplink of a secondary cell are proposed for the allocation of primary channels to the SUs. Spectrum auctions are studied in [20] for energy-efficient channel assignment. Other approaches consider Stackelberg game [22] or contract theory [23] for channel allocation and pricing. However, the limitations of the previous architectures in terms of efficient spectrum use also limit the benefits of spectrum trading. Moreover, the major concern in spectrum trading to date has been spectrum access rather than service delivery. This paper fills this gap by defining joint data and spectrum trading policies in a trusted CDNA which provides operator support for such trading.

CDNA encompasses the advantages of the previous architectures. First, it relieves congestion in the secondary network by offloading traffic through available PUs. Second, it enables efficient use of data and spectrum by reusing available channels temporally and spatially. Third, in contrast to multi-hop cognitive networks, where routes are formed by SUs relaying traffic through primary channels, in our system data is forwarded by PUs through the primary network, which reduces reliability concerns related to multi-hop cognitive transmissions to the first hop (SU-PU link).

## III. NETWORK MODEL

### A. System Architecture for Data and Spectrum Trading

We consider the Cognitive Dynamic Network Architecture (CDNA) shown in Fig. 1 that consists of a PO that incentivizes certain PUs to act as access points for SUs in their vicinity. The SO has its own spectrum bands, although they are potentially congested, and cannot satisfy the QoS requirements of its SUs. Thus, the SO negotiates the conditions for data and spectrum trading with the PO as illustrated in Fig. 1a. Depending on the existing demand, the PO will encourage a set of PUs to share their connectivities with the SUs by offering them a reward. This reward is in return for the PO's benefit obtained when sharing its available data and spectrum with the SO. The SO will allocate the traded channels and PUs to satisfy SU demands. Thus, in CDNA the connectivity is provided by PUs using available channels in the primary network as shown in Fig. 1b. Since the infrastructure is provided by the PO, we omit the SO in the plot. The success of this collaborative market relies on the trustworthiness of the connections which will be elaborated in Section VI.

Suppose that PUs $\mathcal{M} = \{1, 2, …, M\}$ operate in the set of licensed spectrum bands $\mathcal{B} = \{1, 2, …, B\}$, which have an identical bandwidth of size equal to 1. The SUs $\mathcal{N} = \{1, 2, …, N\}$ are equipped with one radio that can be tuned into any available frequency band for packet delivery, i.e., a cognitive radio user can only work on one of the available bands at a time. The availability of the frequency bands varies in time and space as these bands may be occupied by PU transmissions.

The SO harvests licensed spectrum bands, purchases spectrum bands for SU links at different locations, and conducts channel allocation for SU links. The SUs can use the purchased licensed bands and transmit to the available PUs when the primary services are not on, but have to stop using them when primary services become active. This approach is commonly employed in spectrum trading problems to address the interaction between PUs and SUs as a negotiation process [16]. In CDNA, this is enforced by the PO and SO that coordinate the allocation of channels and PUs for the transmission of SUs. Thus, the interference between PUs and SUs can be avoided. The cooperation agreement between the PO and SO relies on resource pricing and revenue sharing to distribute the benefits of this cooperation. This is a reasonable assumption as it might be based on an agreement between companies, in which, in turn, the PO would incentivize the PUs to participate. Since the available resources (data and spectrum) are limited, pricing is used to allocate resources fairly to SUs with respect to their initial purchased data plan. The notation used in the paper is summarized in Table I.

### B. Probability Model for Primary Services

We model the activity of primary services since SU transmission on band $b \in \mathcal{B}$ will depend on the availability of the band. As shown in [24]-[26], primary service traffic can be modeled as a two state ON-OFF process, where ON means that the band is occupied by primary services and OFF means that it is available for opportunistic access by SUs. Let us denote the probability that band $b$ at link $l_{ij}$ is in an OFF state by $a_{ij}^b$ and the probability that band $b$ at link $l_{ij}$ is in an ON state by $(1-a_{ij}^b)$.

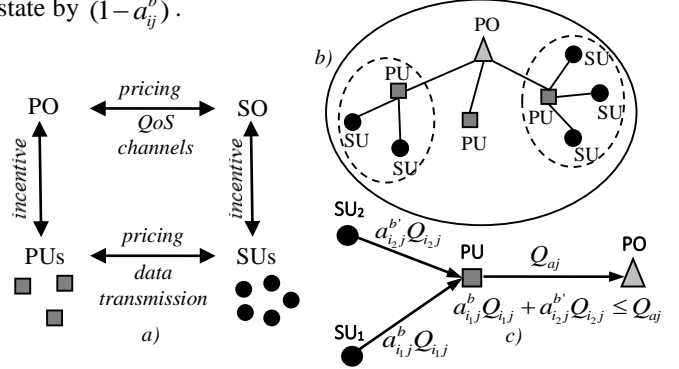

Fig. 1. Cognitive Dynamic Network Architecture
a) Data and Spectrum Trading; b) PU-SU association; c) Data Forwarding

Table I. Notation

| | |
|---|---|
| $N, M, B$ | Total number of SUs, PUs and channels |
| $a_{ij}^b$ | Availability probability of channel $b$ at link $l_{ij}$ |
| $c_{ij}, c_{min,i}$ | Capacity of the link $l_{ij}$, minimum capacity required by SU $i$ |
| $\tau_{ij}, \tau_{min,i}$ | Connectivity / minimum connectivity duration for SU $i$ |
| $Q_{oi}, Q_{oj}$ | Initial data volume for SU $i$ and PU $j$ |
| $x_{ij}^b, t_{ij}^b$ | SU $i$ transmits / trades with PU $j$ in channel $b$ |
| $\rho_{ij}$ | Trustworthiness of link $l_{ij}$ |
| $r_{ij}$ | Reward received by PU $j$ for forwarding the data of SU $i$ |
| $\eta, \sigma, \psi$ | Revenue share in the primary network, secondary network, revenue share among PO and SO |
| $e_j, \xi_j$ | Energy cost of PU $j$, Reliability of PU $j$ |
| $U_S, U_P$ | Utility of SO and PO |
| $U_{ij}^b, V_{ji}^b$ | Utility of SU / PU when SU $i$ transmits to PU $j$ in channel $b$ |
| $\chi, \delta$ | Convergence error / Time index |
| $p_{ij}^b, \varepsilon_{ij}^b, \pi_{ij}$ | Price per unit of data transmitted on channel $b$ in link $l_{ij}$, price per available channel $b$, data price between $i$ and $j$ |
| $O_{ij}^{dir}, O_{ij}^{ind}$ | SU $i$ own direct experience and indirect experience with $j$ |

### C. Link Capacity and Data Volume

Following a widely used model [27], the power propagation gain from SU $i \in \mathcal{N}$ to PU $j \in \mathcal{M}$ is $g_{ij} = \beta \cdot d_{ij}^{-\alpha}$, where $\beta$ is an antenna-related parameter, $\alpha$ is the path loss factor, and $d_{ij}$ is the distance between the two nodes. Let us assume that the transmission power at SU $i$ is $P_i$, and that data transmission is successful only when the received power exceeds a threshold $P_{th}^T$, i.e., $P_i \cdot g_{ij} \geq P_{th}^T$. Thus, we can obtain the transmission range of the SU $i$ as $R_i^T = (\beta \cdot P_i / P_{th}^T)^{1/\alpha}$. Similarly, suppose that the received interference at PU $j$ can be ignored only when its power is less than a threshold $P_{th}^I$. The interference range of PU $j$ can therefore be obtained as $R_j^I = (\beta \cdot P_k / P_{th}^I)^{1/\alpha}$, where $k \in \mathcal{N}$ is an adjacent interfering SU. According to the Shannon-Hartley theorem, if SU $i$ transmits data to PU $j$ using available band $b$, the link capacity will be

$$c_{ij} = \log_2\left(1 + P_i \cdot g_{ij}/\gamma\right) \quad (1)$$

where the bandwidth of band $b$ ($W^b = W = 1$) and $\gamma$ is the Gaussian noise power at PU $j$. In the following section we will address the transmission constraints to avoid interference.

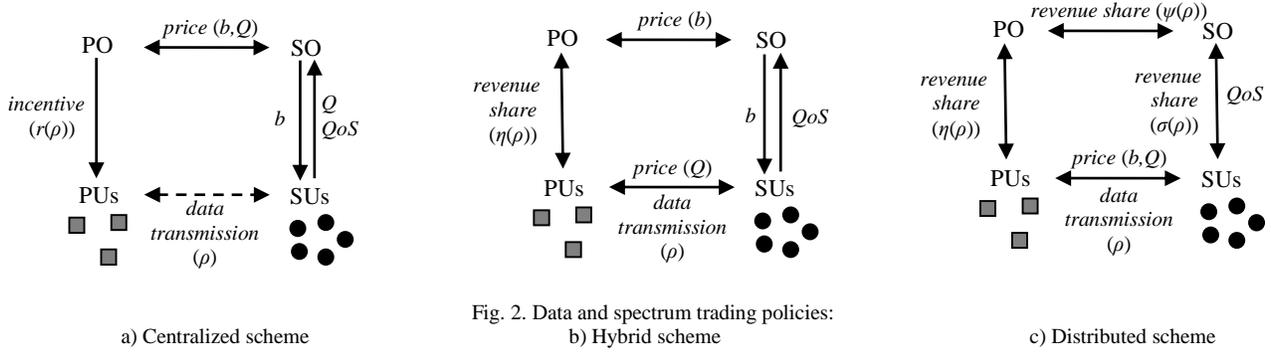

Fig. 2. Data and spectrum trading policies:
a) Centralized scheme  b) Hybrid scheme  c) Distributed scheme

Let us assume that SU $i$ and PU $j$ have a contract with their operator for data volumes $Q_{oi}$ and $Q_{oj}$, respectively. By using the Shannon capacity model, the data volume transmitted between SU $i$ and PU $j$ is denoted as

$$Q_{ij} = c_{ij}\tau_{ij} \quad (2)$$

which varies with the duration of the connection $\tau_{ij}$. The framework for data and spectrum trading is elaborated in the sequel.

## IV. DATA AND SPECTRUM TRADING FRAMEWORK DESIGN

### A. Interference Constraints and Trading Topology

The available bands will be allocated for data transmission to avoid interference between different links. We denote

$$x_{ij}^b = \begin{cases} 1, & \text{if } i \text{ can transmit data to } j \text{ on band } b \\ 0, & \text{otherwise} \end{cases} \quad (3)$$

Let $\mathcal{B}_i \subseteq \mathcal{B}$ denote the set of available licensed bands at SU $i \in \mathcal{N}$. We define the set of PUs in the transmission range of SU $i \in \mathcal{N}$ that can receive in band $b \in \mathcal{B}_i$, as

$$\mathcal{T}_i^b = \{j \mid d_{ij} \leq R_i^T, j \neq i, b \in \mathcal{B}_i\}. \quad (4)$$

Similarly, the set of SUs that can interfere with the reception of PU $j$ on band $b$ is denoted as

$$\mathcal{I}_j^b = \{k \mid d_{kj} \leq R_j^I, k \neq j, b \in \mathcal{B}_k \cap \mathcal{B}_j, \mathcal{T}_k^b \neq \emptyset\} \quad (5)$$

where $\mathcal{B}_k \cap \mathcal{B}_j$ is the set of licensed bands available to SU $k$ and PU $j$, and $\mathcal{T}_k^b \neq \emptyset$ indicates that $k$ has a PU to which it can transmit interfering with reception at $j$.

Based on the previous notations, we present the interference constraints. An SU $i \in \mathcal{N}$ will be associated with a PU $j \in \mathcal{M}$ on at most one band $b$ at a time, and a PU $j \in \mathcal{M}$ cannot receive from multiple SUs on the same band,

$$\text{I1}: \sum_b \sum_{j \in \mathcal{T}_i^b} x_{ij}^b \leq 1 \text{ and } \sum_{\{i \mid j \in \mathcal{T}_i^b\}} x_{ij}^b \leq 1. \quad (6)$$

Interference between adjacent SUs must also be considered. According to (4), we note that when SU $i \in \mathcal{N}$ is transmitting data on band $b \in \mathcal{B}_i$, any other SU that can interfere with the reception of PU $j$ cannot use this band. Thus, we have the following constraint,

$$\text{I2}: x_{ij}^b + \sum_{\{q \in \mathcal{T}_k^b\}} x_{kq}^b \leq 1, \quad k \in \mathcal{I}_j^b, k \neq i. \quad (7)$$

Let us denote by $t_{ij}^b$ the association between SU $i \in \mathcal{N}$ and PU $j \in \mathcal{M}$ on a particular channel $b \in \mathcal{B}$ for data and spectrum trading purposes with

$$t_{ij}^b = \begin{cases} 1, & \text{if } i \text{ associates with } j \text{ on band } b \\ 0, & \text{otherwise} \end{cases} \quad (8)$$

The trading association must satisfy the constraint, $t_{ij}^b \leq x_{ij}^b$. The *trading topology* $\mathbf{T}=[t_{ij}^b]$ provides information on the existing trading associations throughout the network. We assume that separate channels are used for the transmission of control messages between SUs and PUs for the association process. Thus, these messages do not interfere with data transmission.

### B. Problem Formulation

Data and spectrum trading involves the PO, SO and their end users. Besides, each SU $i$ has minimum requirements in terms of capacity $c_{min,i}$ and availability of service duration $\tau_{min,i}$ that a PU $j$ must satisfy,

$$a_{ij}^b c_{ij} \geq c_{\min,i} \quad (9)$$

$$\tau_{ij} \geq \tau_{\min,i} \quad (10)$$

The previous requirements result in the following demand for data volume from the SU: $Q_{\min,i} = c_{\min,i}\tau_{\min,i}$.

In this architecture, transmission from the SU to the base station (BS) is conducted in two hops by a decode and forward scheme: SU→PU and PU→BS. Let $Q_{aj}$ denote the available data of PU $j$. Then, the data volume transmitted between SU $i$ and PU $j$ is $a_{ij}^b Q_{ij} \leq Q_{aj}$, where $a_{ij}^b$ is the channel availability in link $l_{ij}$ and $Q_{ij}$ is given by (2). Based on the requirements of SU $i$ in the first hop (9)-(10) and transmission availability in the second hop, the PU will decide whether it accepts or not the association. After the PU relays the data from the SU, its data volume is reduced to $Q_{oj} - a_{ij}^b Q_{ij}$ and the data of SU $i$ is extended to $Q_{oi} + a_{ij}^b Q_{ij}$. The relaying process is illustrated in Fig. 1c when SUs $i_1$ and $i_2$ get associated to PU $j$. If PU $j$ does not transmit the agreed-on data its reliability will decrease and so will the trustworthiness of the connection. Second link transmissions will be scheduled by the primary operator (PO) and their analysis is beyond the scope of this paper.

In the following subsections we present different policies for data and spectrum trading based on centralized, hybrid and distributed resource allocation. These policies rely on different revenue-sharing schemes between the PO and SO ($\psi$), the PO and PUs ($\eta$), and the SO and SUs ($\sigma$) as illustrated in Fig. 2. In these schemes (from left to right), the PO and SO delegate trading progressively to their end users and adopt more flexible trading polices to reduce complexity. The association between SUs and PUs depends on the trustworthiness of the connection $\rho$, which will also impact the level of revenue share. For clarity of presentation, first we design each trading framework for a given trustworthiness $\rho$ and revenue share $\psi(\rho)=\psi$, $\eta(\rho)=\eta$ and $\sigma(\rho)=\sigma$, and then we elaborate in detail the trust relationships and their influence on the revenue share in Section VI. By *backward induction*, we analyze the strategies of the different parties involved in the trading for each scheme.

*B1. Centralized Scheme*

Based on the SUs' demand for data volume and QoS requirements, the SO identifies the necessary channels and available PUs to satisfy the demand. It then negotiates the price for these resources with the PO. Once both parties agree, the SO assigns to each SU the channel and PU that satisfies its QoS requirements. Finally, the PO rewards the PU for sharing its connectivity. In this scheme, as illustrated in Fig. 2a, the service is guaranteed by the operators through pricing.

Suppose that the price per unit of data transmitted on channel $b \in \mathcal{B}$ in link $l_{ij}$ is $p_{ij}^b$. The utility of the SO is defined as the difference between the social welfare in the secondary network and the price paid to the PO for the resources,

$$U_S = \sum_{i=1}^{N} \sum_{j=1}^{M} \sum_{b=1}^{B} U_{ij}^b - t_{ij}^b p_{ij}^b a_{ij}^b Q_{ij} \quad (11)$$

where $U_{ij}^b$ is the utility of the SU $i$ transmitting to PU $j$ on channel $b$. It provides the value SU $i$ gives to the initial data volume $Q_{oi}$ and demanded data volume $Q_{ij}$,

$$U_{ij}^b = t_{ij}^b \rho_{ij} \log(Q_{oi} + a_{ij}^b Q_{ij}) \quad (12)$$

and $0 < \rho_{ij} \le 1$ is the trustworthiness of the connection. Its calculation is elaborated on Section VI. Pricing is used to allocate the resources fairly to SUs with respect to their initial purchased data. Users who have purchased a higher amount of initial data value the resources more and, thus, they will have more chances to get additional resources.

Given the price $p_{ij}^b$ announced by the PO, the centralized optimization problem at the SO is as in (13), where $t_{ij}^b$ provides the channel allocation and SU-PU association and $Q_{aj}$ is the available data at PU $j$. This optimization yields the optimum trading topology $\mathbf{T}^* = [(t_{ij}^b)^*]$ and data traded $Q_{ij}^*$.

$$\underset{t_{ij}^b, Q_{ij}}{\text{maximize}} \quad U_S = \sum_{i=1}^{N} \sum_{j=1}^{M} \sum_{b=1}^{B} t_{ij}^b (\rho_{ij} \log(Q_{oi} + a_{ij}^b Q_{ij}) - p_{ij}^b a_{ij}^b Q_{ij})$$

subject to $\quad t_{ij}^b \le x_{ij}^b, \ x_{ij}^b \in \{0,1\}, i \in \mathcal{N}, j \in \mathcal{T}_i^b, b \in \mathcal{B}_i \cap \mathcal{B}_j$

(6), (7), (9), (10)

$$Q_{\min,i} \le a_{ij}^b Q_{ij} \le Q_{aj} \quad (13)$$

The SO will negotiate with the PO the price for the data and channels needed to satisfy the SUs' optimal demand. This optimization problem is a mixed-integer linear programming problem with complexity $\mathcal{O}(N^{MB})$. It can be optimally solved with standard algorithms (e.g., sequential fixing algorithm [29], branch and bound [30]) or software (e.g., CPLEX [31] and MATLAB) for small scenarios.

Let $r_{ij}$ denote the incentive in monetary units to compensate PU $j$ for forwarding the data of SU $i$. The incentive is defined as $r_{ij} = (t_{ij}^b)^* \eta \Phi_{oj} a_{ij}^b Q_{ij}^* / Q_{aj}$, where $\eta$ is a revenue share between PO and PU, $\Phi_{oj}$ is the price of the data plan of PU $j$ and $Q_{aj}$ is the available data at PU $j$. Physically, this means that the overall data forwarded will be compensated with an amount proportional to the price of the remaining available data. The utility of the PU is

$$V_{ji}^b = (t_{ij}^b)^* \rho_{ji} \log(Q_{oj} - a_{ij}^b Q_{ij}^*) + r_{ij} - e_j \quad (14)$$

where $\rho_{ji}$ is the trust of PU $j$ in the connection requested by SU $i$, and $e_j$ is the energy cost for serving as an access point. The details on the trust relationships are provided in Section VI.

Next, the utility of the PO is defined as the difference between the payment from the SO and the incentive offered to each PU,

$$U_P = \sum_{i=1}^{N} \sum_{j=1}^{M} \sum_{b=1}^{B} (t_{ij}^b)^* p_{ij}^b a_{ij}^b Q_{ij}^* - r_{ij} \quad (15)$$

The PO and SO negotiate the price $p_{ij}^b$ iteratively and in parallel for each SU $i \in \mathcal{N}$ and PU $j \in \mathcal{M}$ in channel $b \in \mathcal{B}$ as outlined in Algorithm 1. In each iteration of the negotiation process a new association $(t_{ij}^b)^*$ and $Q_{ij}^*$ are obtained solving (13) for the new negotiated price. After successive negotiations an agreement is reached[2] and the agreed price is $(p_{ij}^b)^*$. The convergence speed of the algorithm can be controlled by tuning the values of the convergence error $\chi$ and $\Delta p_{ij}^b$. A fine grained adjustment of $\Delta p_{ij}^b$ will result into smaller $\chi$ but slower convergence. The detailed proof of convergence is shown in the Appendix. A similar collaborative negotiation process is used in [36] to quantify the incentives for cooperation between cellular and small service operators. The negotiation mechanism leads to fair sharing of the benefits in each joint access network decision. For a comprehensive survey on pricing theory for spectrum trading, see [37].

---

[2]Let us recall that the cooperation between the PO and SO relies on resource pricing and revenue sharing to distribute the benefits of cooperation among the parties involved. The condition to reach an agreement is $|U_S - \alpha \cdot U_P| \le \chi$, where $\alpha$ indicates the sharing of benefits between PO and SO, and $\chi$ is the convergence error. For simplicity, we have assumed the benefits are shared equally (i.e., $\alpha = 1$). This requirement is used as a stop criterion in the algorithms presented in the paper. Nevertheless, our scheme admits any other sharing of benefits through parameter $\alpha$.

Assuming that $I_1$ iterations are needed to reach the agreement, the complexity of this algorithm is $\mathcal{O}(I_1(N^{MB}+NM + NMB))$, where the first term inside the inner parentheses is the complexity of solving problem (13), the second term is the complexity of calculating the PU reward for serving each SU, and the third term is the complexity of calculating the PO utility.

The centralized scheme is provided as a benchmark. In the next section, the SO and PO will progressively delegate the trading to the SUs and PUs to reduce the complexity of the problem but still receive some revenue.

---

**Algorithm 1** Centralized Data and Spectrum Trading

1: PO announces an initial price $p^b_{ij}$ and a revenue share $\eta$ with PUs
2: SO solves (13) for current demand and price $p^b_{ij}$ to obtain $(\mathbf{T}^b)^*$ and $Q^*_{ij}$
3: PO calculates the revenue $r_{ij}$ for PU $j$ for forwarding data $Q^*_{ij}$
4: PO calculates its utility in (15)
5: SO and PO negotiate the price $p^b_{ij}$ as follows:
6: **If** $U_S > U_P + \chi$
7:     PO announces a new price $p^b_{ij} \leftarrow p^b_{ij} + \Delta p^b_{ij}$
8:     Go back to 2)
9: **elseif** $U_S < U_P - \chi$
10:    PO announces a new price $p^b_{ij} \leftarrow p^b_{ij} - \Delta p^b_{ij}$
11:    Go back to 2)
12: **elseif** $|U_S - U_P| \leq \chi$
13:    The optimum association $(\mathbf{T}^b)^*$, data $Q^*_{ij}$ and trading price for the data and channels $(p^b_{ij})^*$ are obtained
14: **end**

---

*B2. Hybrid Scheme*

In the hybrid scheme, the SO and PO negotiate the price for the channels while the SUs and PUs negotiate the price for the data. The PO allows PUs to trade with their data and benefit from the transaction. Each SU selects the most convenient PU, i.e., the PU with sufficient data availability and service duration to meet its QoS requirements. After the price and amount of data traded are agreed on, the SO determines which channels are needed for each SU-PU link to meet these requirements. The SO then negotiates the corresponding price for these channels with the PO and allocates them to each SU-PU link for data transmission. Finally, the PU pays the PO a percentage of the revenue earned in trading the data. This scheme is illustrated in Fig. 2b.

By abusing the notation, let us denote by $t_{ij} = 1$ when there is a trade between $i$ and $j$ ($Q_{ij} > 0$), and $t_{ij} = 0$ otherwise. The utility of the SU is redefined to include the data trading price $\pi_{ij}$ between SU $i$ and PU $j$ as,

$$U_{ij} = t_{ij}(\rho_{ij}\log(Q_{oi}+Q_{ij}) - \pi_{ij}Q_{ij}) \quad (16)$$

where $Q_{ij}$ is the data traded. Notice that the data traded does not consider the channel that will be later assigned by the SO. The PU will sell its data volume in return for profit. This profit can be defined as the gain from serving an SU and the price charged for selling the data. Thus, the utility of a PU when selling data volume $Q_{ji}$ is

$$V_{ji} = t_{ij}(\rho_{ji}\log(Q_{oj}-Q_{ji}) + \eta\pi_{ij}Q_{ji} - e_j) \quad (17)$$

where $\eta$ is the revenue share between the PO and the PU.

Considering the connectivity requirements of the SU and the price of the data $\pi_{ij}$, the SU optimization is as follows,

$$\underset{t_{ij},Q_{ij}}{\text{maximize}} \quad U_{ij} = t_{ij}(\rho_{ij}\log(Q_{oi}+Q_{ij}) - \pi_{ij}Q_{ij})$$

subject to $\quad Q_{\min,i} \leq Q_{ij} \leq Q_{aji}$, (10) $\quad (18)$

where $Q_{\min,i} = c_{\min,i}\tau_{\min,i}$ and $Q_{aji}$ is the available data PU $j$ is willing to share with SU $i$. In response to demand $Q^*_{ij}$, we can then formulate the PU optimization problem as

$$\underset{Q_{ji}}{\text{maximize}} \quad V_{ji} = t^*_{ij}(\rho_{ji}\log(Q_{oj}-Q_{ji}) + \eta\pi_{ji}Q_{ji} - e_j)$$

subject to $\quad \sum_i Q_{ji} \leq Q_{aj}, \; Q_{\min,i} \leq Q_{ji} \leq Q^*_{ij}$ $\quad (19)$

where the first constraint indicates that the total data volume sold by PU $j$ cannot exceed its available data volume, and the second constraint guarantees that the data sought satisfies the demand.

Optimization problems (18) and (19) are solved iteratively until the optimum values $t^*_{ij}$ and $Q^*_{ij}$ are obtained as detailed in Algorithm 2, together with the negotiation of the data price $\pi^*_{ij}$. The utility in (18) decreases linearly with $\pi_{ij}$ and in (19) increases linearly with respect to the same variable. Thus, they intersect at a single point $\pi^*_{ij}$. Then, the SO assigns the channels to maximize its utility which is defined as

$$U_S = \sum_{i=1}^{N}\sum_{j=1}^{M}\sum_{b=1}^{B} U^b_{ij} - t^b_{ij}\varepsilon^b_{ij}a^b_{ij} \quad (20)$$

where $\varepsilon^b_{ij}$ is the price per available channel charged by the PO. The SO optimization problem is as follows

$$\underset{t^b_{ij}}{\text{maximize}} \quad U_S = \sum_{i=1}^{N}\sum_{j=1}^{M}\sum_{b=1}^{B} t^b_{ij}(\rho_{ij}\log(Q_{oi}+a^b_{ij}Q^*_{ij}) - a^b_{ij}(Q^*_{ij}\pi^*_{ij}+\varepsilon^b_{ij}))$$

subject to $\quad t^b_{ij} \leftarrow t^*_{ij}, b, \; \forall b \in \mathcal{B}_i \cap \mathcal{B}_j$

$\quad t^b_{ij} \leq x^b_{ij}, \; (6),(7)$

$\quad Q^*_{ij} \leq a^b_{ij}Q^*_{ij}$ $\quad (21)$

The frequency bands will be reused by SUs to reduce costs and assigned to the most profitable links to satisfy the negotiated data needs. This optimization problem is an integer linear programming problem and can be optimally solved in polynomial time using standard algorithms (e.g., sequential fixing algorithm [29], branch and bound [30]) or software (e.g., CPLEX [31] and MATLAB).

Finally, the PO calculates its utility as

$$U_P = \sum_{i=1}^{N}\sum_{j=1}^{M}\sum_{b=1}^{B}(t^b_{ij})^*(1-\eta)\pi^*_{ij}a^b_{ij}Q^*_{ij} + (t^b_{ij})^*a^b_{ij}\varepsilon^b_{ij} \quad (22)$$

where $\eta$ is the revenue share between the PO and the PU.

The PO and SO iteratively negotiate the price per available channel and the optimum value $(\varepsilon^b_{ij})^*$ is obtained when $|U_S - U_P| \leq \chi$. The selection of the step price $\Delta\varepsilon^b_{ij}$ needed for the convergence of the price negotiation follows the same

reasoning as the proof of convergence for Algorithm 1 in the Appendix. Due to space limitations the details are omitted.

In this scheme, the price charged to the SO and SUs per channel and unit of data traded is $p_{ij}^b = \pi_{ij} + \varepsilon_{ij}^b / Q_{ij}$, and the incentive that PU $j$ receives for sharing its resources is $r_{ij} = t_{ij}^* \eta \pi_{ij}^* Q_{ij}^*$. Since the channel is allocated to SU-PU links after the data negotiation, the uncertainty surrounding the final amount of data transmitted is higher than in the centralized approach. To compensate for this uncertainty, the price should be reduced accordingly as $\pi_{ij} \leq \Phi_{oj} / Q_{aj}$.

This scheme is outlined in Algorithm 2. Let us denote by $I_2$ the number of iterations needed to solve the SU and PU optimization problems, and by $I_2'$ the number of iterations for the PO and the SO to reach an agreement. Then, the complexity of this algorithm is $\mathcal{O}(I_2(N^M + NM) + I_2'(NM^B + NMB))$, where the first term is the complexity of solving (18) and (19), respectively, and the second term is the complexity of solving (21) and (22), respectively. The complexity of Algorithm 2 is significantly reduced compared to Algorithm 1.

---

**Algorithm 2** Hybrid Data and Spectrum Trading
1: PU sets an initial data price $\pi_{ij}$ and $Q_{aji} = Q_{aj}$
2: SU solves (18) for the current price $\pi_{ij}$ to obtain $\mathbf{T}^*$ and $Q^*_{ij}$
3: PU solves (19) for the previous demand and obtains $Q^*_{ji}$
4: Update $Q_{aji} = Q^*_{ji}$
5: Go to 2) until $Q^*_{ij} = Q^*_{ji}$
6: PU provides a new price $\pi_{ij}$ as follows:
4: **If** $U_{ij} > V_{ji} + \chi'$
5:   PU announces a new price $\pi_{ij} \leftarrow \pi_{ij} + \Delta \pi_{ij}$
6:   Go back to 2)
7: **elseif** $U_{ij} < V_{ji} - \chi'$
8:   PU announces a new price $\pi_{ij} \leftarrow \pi_{ij} - \Delta \pi_{ij}$
9:   Go back to 2)
10: **elseif** $|U_{ij} - V_{ji}| \leq \chi'$
11:   The optimum price $\pi^*_{ij}$ and data $Q^*_{ij}$ are obtained
12: **end**
13: PO announces the channel price $\varepsilon^b_{ij}$
14: SO solves (21) for channel price $\varepsilon^b_{ij}$ to obtain the optimum channel allocation $(\mathbf{T}^b)^*$
15: SO and PO negotiate the price $\varepsilon^b_{ij}$ as follows:
16: **If** $U_S > U_P + \chi$
17:   PO announces a new price $\varepsilon^b_{ij} \leftarrow \varepsilon^b_{ij} + \Delta \varepsilon^b_{ij}$
18:   Go back to 10)
19: **elseif** $U_S < U_P - \chi$
20:   PO announces a new price $\varepsilon^b_{ij} \leftarrow \varepsilon^b_{ij} - \Delta \varepsilon^b_{ij}$
21:   Go back to 10)
22: **elseif** $|U_S - U_P| \leq \chi$
23:   The optimum association $(\mathbf{T}^b)^*$, data $Q^*_{ij}$ and trading price for the data and channels $(\varepsilon^b_{ij})^*$ are obtained
24: **end**

---

*B3. Distributed Scheme*

In the distributed scheme, the PO and SO delegate data and spectrum trading to their respective users. In other words, SUs negotiate the price for trading data and channels with PUs. The PO and SO agree on an initial revenue share for their cooperation and assist their users in the association process by providing information on the available channels. They both expect to obtain additional benefits from the trade. This scheme is shown in Fig. 2c.

Let us assume that $b \in \mathcal{B}_i \cap \mathcal{B}_j$. The utility of the SU and PU are redefined, respectively, as,

$$U_{ij}^b = t_{ij}^b (\rho_{ij} \log(Q_{oi} + a_{ij}^b Q_{ij}) - (1-\sigma) p_{ij}^b a_{ij}^b Q_{ij}) \quad (23)$$

$$V_{ji}^b = t_{ij}^b (\rho_{ji} \log(Q_{oj} - a_{ji}^b Q_{ji}) + \eta p_{ij}^b a_{ij}^b Q_{ji} - e_j) \quad (24)$$

where $p_{ij}^b$ is the price per channel and unit of data transmitted and $\sigma$ is the revenue share between the SO and SU (the SO compensates the SU by paying a percentage of the data cost for transmitting in the primary network). The optimization problem for each SU is formulated as:

$$\underset{t_{ij}^b, Q_{ij}}{\text{maximize}} \quad U_{ij}^b = t_{ij}^b (\rho_{ij} \log(Q_{oi} + a_{ij}^b Q_{ij}) - (1-\sigma) p_{ij}^b a_{ij}^b Q_{ij}) \quad (25)$$

subject to $\quad Q_{\min,i} \leq Q_{ij} \leq Q_{aji}, (10)$

And the optimization problem for each PU is formulated as:

$$\underset{p_{ij}^b, Q_{ji}}{\text{maximize}} \quad V_{ji}^b = (t_{ij}^b)^* (\rho_{ji} \log(Q_{oj} - a_{ji}^b Q_{ji}) + \eta p_{ij}^b a_{ij}^b Q_{ji} - e_j)$$

subject to $\quad \sum_i Q_{ji} \leq Q_{aj}, \; Q_{\min,i} \leq Q_{ji} \leq Q_{ij}^* \quad (26)$

To solve the previous optimization problems distributively and with low computational complexity, we develop a *distributed data and spectrum trading* algorithm (Algorithm 3), described in the next section, based on matching theory. This algorithm provides the optimum $(t_{ij}^b)^*$, $Q_{ij}^*$ and $(p_{ij}^b)^*$.

Let us assume that the PO and SO agree on a revenue share $\psi$. Then, the utility of the SO is

$$U_S = \sum_{i=1}^N \sum_{j=1}^M \sum_{b=1}^B (1-\psi)[U_{ij}^b - (t_{ij}^b)^* \sigma (p_{ij}^b)^* a_{ij}^b Q_{ij}^*]$$

where $\sigma$ is the percentage of the price granted to the SU.

Accordingly, the utility of the PO is given by the revenue share with the SO $\psi$ and the percentage $1 - \eta$ of the gain obtained by the PU for trading its data,

$$U_P = \sum_{i=1}^N \sum_{j=1}^M \sum_{b=1}^B \psi [U_{ij}^b - (t_{ij}^b)^* \sigma (p_{ij}^b)^* a_{ij}^b Q_{ij}^*] + (t_{ij}^b)^* (1-\eta)(p_{ij}^b)^* a_{ij}^b Q_{ij}^*$$

The optimum revenue share $\psi^*$ is obtained iteratively when $|U_S - U_P| \leq \chi$. The revenue-sharing policies for the three schemes are shown in Table II, and their complexity is summarized in Table III.

Table II. Revenue sharing policies

|  | Centralized | Hybrid | Distributed / Matching |
|---|---|---|---|
| PO → PU | $\eta \Phi_{oj} a_{ij}^b Q_{ij} / Q_{aj}$ | $\eta \pi_{ij} Q_{ij}$ | $\eta p_{ij}^b a_{ij}^b Q_{ij}$ |
| SO → PO | $p_{ij}^b a_{ij}^b Q_{ij}$ | $a_{ij}^b \varepsilon_{ij}^b$ | $\psi(U_{ij}^b - \sigma p_{ij}^b a_{ij}^b Q_{ij})$ |
| SU → PU | — | $\pi_{ij} Q_{ij}$ | $p_{ij}^b a_{ij}^b Q_{ij}$ |
| SO → SU | — | — | $\sigma p_{ij}^b a_{ij}^b Q_{ij}$ |

Table III. Computational complexity

| Centralized (Algorithm 1) | Hybrid (Algorithm 2) | Distributed / Matching (Algorithm 3) |
|---|---|---|
| $\mathcal{O}(I_1(N^{MB}+NM+NMB))$ | $\mathcal{O}(I_2(N^M+NM) + I_2'(NM^B+NMB))$ | $\mathcal{O}(NMBI_3(\log(MB) + \log(NB))$ |

## V. JOINT DATA AND SPECTRUM TRADING MATCHING ALGORITHM

Our next goal is solving the distributed data and spectrum trading problem (25)-(26) with low computational complexity to capture network dynamics. In a dense CDNA it is desirable to solve resource allocation in a decentralized and self-organized way to facilitate SU-PU associations and data and spectrum trading decisions without having to rely on a centralized controller. One suitable tool for developing decentralized and self-organized solutions, which can solve the optimization problem and avoid combinatorial complexity, is the framework of matching games [32]. In this regard, we consider matching theory to solve the distributed SU and PU optimization problems in (25)-(26), which involve SU-PU associations, channel allocation and price for data and channels as explained in Section IV.B3. In a matching game two sets of players must be assigned to each other according to their preferences. Each player ranks the players in the other set using a preference relation. Therefore, we formulate the problem as a two-sided many-to-one matching game in which each SU $i \in \mathcal{N}$ can be associated with only one PU $j \in \mathcal{M}$ that satisfies its QoS requirements. In addition, each PU can admit a certain quota of users served on different channels ($\sum_i \sum_b a_{ij}^b Q_{ji} \leq Q_{aj}$). SUs and PUs rank one another based on their respective utilities defined in (23) and (24). In the sequel, we develop the matching algorithm to solve the data and spectrum trading distributively for a snap shot of the network, and in Section VII we present the dynamic tracking.

### A. Data and Spectrum Trading Game

We formally define the *data and spectrum trading game* by the tuple $(\mathcal{N}, \mathcal{M}, \mathcal{B}, \succ_\mathcal{N}, \succ_\mathcal{M})$. Here, $\succ_\mathcal{N} = \{\succ_i\}_{i \in \mathcal{N}}$ and $\succ_\mathcal{M} = \{\succ_j\}_{j \in \mathcal{M}}$ denote, respectively, the set of preference relations of SUs and PUs. For any two PUs $j, j' \in \mathcal{M}$ and two channels $b, b' \in \mathcal{B}$, the preference relation $\succ_i$ for SU $i$ according to (23) is defined as: $(j,b) \succ_i (j,b') \Leftrightarrow U_{ij}^b > U_{ij}^{b'}$ and $(j,b) \succ_i (j',b) \Leftrightarrow U_{ij}^b > U_{ij'}^b$. Similarly, the preference relation $\succ_j$ for PU $j$ over two SUs $i, i' \in \mathcal{N}$, and two channels $b, b' \in \mathcal{B}$ according to (24) is defined as: $(i,b) \succ_j (i,b') \Leftrightarrow V_{ji}^b > V_{ji}^{b'}$ and $(i,b) \succ_j (i',b) \Leftrightarrow V_{ji}^b > V_{ji'}^b$.

**Definition 1.** A many-to-one matching problem in CDNA is a function $\mu$ from the set $\mathcal{N} \cup \mathcal{M} \cup \mathcal{B}$ into the set $\mathcal{N} \cup \mathcal{M} \cup \mathcal{B}$ such that
1) $j = \mu(i,b)$ if and only if $i = \mu(j,b)$,
2) $|\mu(i,b)| = 1$ and $|\mu(j,b)| = 1$,
3) $\sum_b |\mu(i,b)| = 1$ and $\sum_b |\mu(j,b)| \leq \min(B, n_{aj})$ where $B$ is the number of channels and $n_{aj}$ is the number of SUs that can be served by PU $j$ with $Q_{aj}$.
4) $|\mu(i,b)| + \sum_{k \in \mathcal{I}_{\mu(i,b)}^b} |\mu(k,b)| \leq 1$ where $\mathcal{I}_{\mu(i,b)}^b$ is the set of SUs that can interfere with the reception of $\mu(i,b)$ on band $b$.

**Definition 2.** A pair $(i, j) \notin \mu$ where $i \in \mathcal{N}$, $j \in \mathcal{M}$ is a *blocking pair* for the matching $\mu$, if (i) $(j,b) \succ_i (\mu(i,b), b)$ or $(j,b) \succ_i (\mu(i,b'), b')$, **and** (ii) $(i,b) \succ_j (\mu(j,b), b)$ or $(i,b) \succ_j (\mu(j,b'), b')$. Accordingly, a matching is blocked by $(i, j)$ when SU $i$ and PU $j$ prefer each other to their current matching on the same or different channel. A matching $\mu$ is *stable* if and only if there is no blocking pair.

A matching solution for the data and spectrum trading game is stable if after the SU and PU agree on the amount of data and price for transmission on a particular channel, there are no SU-PU associations or channel allocations that will improve the current matching. In other words, the utility of the SU and PU in (23) and (24) will not be increased by if another PU or SU or another channel is selected.

### B. Distributed Data and Spectrum Trading Algorithm

To solve the formulated *data and spectrum trading game*, we propose a novel distributed algorithm (Algorithm 3) that allows the players to self-organize into a stable matching that guarantees their connectivity requirements. The proposed algorithm consists of two main stages: Stage 1 focuses on matching and price initialization and Stage 2 determines the matching and trading price.

First, we assume an initial price for the data $p_{ij}^b(0)$. This will be updated later on based on data demand and supply. In Stage I, each SU $i$ selects for every available channel $b$ a set of PUs that satisfy its QoS requirements, denoted by set $\mathcal{M}_i^b$, and sorts them in decreasing order according to the utility function in (23). Similarly, each PU $j$ selects a set of SUs denoted by $\mathcal{N}_j^b$ in decreasing order according to the utility in (24).

The data demand and supply functions for each potential SU-PU association are obtained by differentiating the utilities in (23) and (24) with respect to $Q_{ij}$ and $Q_{ji}$ and making them equal to zero [33], respectively,

$$\mathcal{D}_{ij}^b = Q_{ij} = \frac{1}{a_{ij}^b}\left(\frac{\rho_{ij}}{p_{ij}^b(1-\sigma)} - Q_{oi}\right) \qquad (27)$$

$$\mathcal{S}_{ji}^b = Q_{ji} = \frac{1}{a_{ij}^b}\left(Q_{oj} - \frac{\rho_{ji}}{p_{ij}^b \eta}\right) \qquad (28)$$

The demand and supply functions depend on the channel availability $a_{ij}^b$ at link $l_{ij}$, the trustworthiness of the connection $\rho_{ij}$, the data price $p_{ij}^b$, the price compensation by the SO $\sigma$ and by the PO $\eta$, respectively, and the initial amount of data purchased $Q_{oi}$ and $Q_{oj}$, respectively. The calculation of the trustworthiness is elaborated in Section VI. SUs and PUs exchange their demand and supply per channel $b$ with their selected counterparts to accurately estimate the price. The total data demand and supply on channel $b$ at equilibrium is $\sum_{i \in \mathcal{N}_j^b} \sum_{j \in \mathcal{M}_i^b} \mathcal{D}_{ij}^b = \sum_{i \in \mathcal{N}_j^b} \sum_{j \in \mathcal{M}_i^b} \mathcal{S}_{ji}^b$ and, the equilibrium price is derived iteratively using the following equation,

$$p_{ij}^b(\delta+1) = p_{ij}^b(\delta) + \varpi_j \left( \sum_{i \in \mathcal{N}_j^b} \sum_{j \in \mathcal{M}_i^b} \mathcal{D}_{ij}^b(\delta) - \mathcal{S}_{ji}^b(\delta) \right) \quad (29)$$

where the price in the next iteration is the difference between demand and supply at time $\delta$, weighted by the learning rate $\varpi$ and added to the price in the current iteration. This pricing function adjusts the demand and supply until the price converges to the optimal price. The stability of the price depends on the learning rate $\varpi_j$ which is analyzed in Section V.C.

Next, each SU and PU updates the utility according to (23) and (24), respectively, with the new price $p_{ij}^b$ and re-orders its lists of preferences. After the initialization, the matching is formed in Stage II. Each SU $i$ sends a proposal to its preferred PUs as in $\mathcal{M}_i^b$. The PUs accept or reject the proposals according to their preferences as per $\mathcal{N}_j^b$. This process is repeated until all SUs are matched to a PU or rejected by all their preferred PUs. The new price is obtained by (29) for the current matching with $\mathcal{M}_i^b(\mu)$ and $\mathcal{N}_j^b(\mu)$. The PUs then broadcast the new price $p_{ij}^b$, which includes information on the allocated channels and selected SUs so that adjacent PUs will not allocate the same channels if there is interference. SU and PU preferences may change with the new matching and, if this happens, they will need to update their list of preferences. The algorithm terminates once $\mathcal{M}_i^b(\mu)$ and $\mathcal{N}_j^b(\mu)$ remain the same for two consecutive matchings. As a result, the trading topology $(t_{ij}^b)^* = \{1 \mid i \in \mathcal{N}_j^b(\mu), j \in \mathcal{M}_i^b(\mu)\}$ is obtained together with the optimal price $(p_{ij}^b)^*$ as in (29), for the optimum data transmitted per unit of spectrum $(Q_{ij})^*$ obtained by (27)-(28) at the equilibrium.

---

**Algorithm 3** Distributed Data and Spectrum Trading

1: **Procedure – Price calculation**
2: **for** each SU $i$ :
3:   **for** each $j \in \mathcal{M}_i^b$ :
4:     Obtain local data demand (27), supply (28), and learning rate (30)
5:     Obtain price $p^b{}_{ij}(\delta+1)$ using (29)
6:     **while** $|p^b{}_{ij}(\delta+1) - p^b{}_{ij}(\delta)| > \varepsilon$ **do**
7:       Update data demand (27) and supply (28)
8:       Calculate learning rate $\varpi_j$ (30) and update price $p^b{}_{ij}(\delta+1)$ using (29)
9:       $\delta = \delta + 1$
10:     **end**
11:   **end**
12: **end**
13: **Stage 1 – Initialization**
14: Initialize the price to $p^b{}_{ij}(0)$
15: Each SU $i$ chooses a set of PUs on channel $b$ $\mathcal{M}_i^b$ following $\succ_i$ as in (23)
16: Each PU $j$ selects a set of SUs $\mathcal{N}_j^b$ following $\succ_j$ as in (24)
17: Obtain the price $p^b{}_{ij}(\delta)$ for the initial demand and supply
18: Obtain $U_{ij}^b$ and $V_{ji}^b$ with the new price and reorder $\mathcal{M}_i^b$ and $\mathcal{N}_j^b$
19: **Stage 2 – Matching and price determination**
20: Each SU $i$ issues proposals to its preferred PU, and the PU accepts or rejects the proposal
21: Calculate price (29) for current matching $\mu$ with $\mathcal{M}_j^b(\mu)$ and $\mathcal{N}_j^b(\mu)$
22: Obtain $U_{ij}^b$ and $V_{ji}^b$ with the new price and reorder $\mathcal{M}_i^b$ and $\mathcal{N}_j^b$
23: Go back to 20 until $\mathcal{M}_j^b(\mu)$ and $\mathcal{N}_j^b(\mu)$ remain unchanged for two consecutive iterations
24: As result, $(t_{ij}^b)^*$, $(p_{ij}^b)^*$ and $(Q_{ij})^*$ are obtained

---

*C. Stability of the Proposed Algorithm*

Let us first analyze the externalities of the data and spectrum market.

**Remark 1.** The proposed data and spectrum trading game has positive and negative externalities.

Indeed, the preferences of SUs and PUs change as the game evolves. In CDNA, the more PUs participating in the market, the more likely an SU will find a good match and vice versa. We study the externalities by focusing on the local network effects of SUs and PUs (local CDNA).

For an SU $i \in \mathcal{N}_j^b(\mu)$:

- A new available PU $j'$ is a *positive externality* if it increases $U_{ij}^b$, $j \in \mathcal{M}_i^b(\mu)$ due to any of the following reasons: a) $\exists i'$, $i' \in \mathcal{N}_j^b(\mu) / (j',b) \succ_{i'} \mu(j,b)$, hence the quality of the connection between SU $i$ and PU $j$ improves, or, b) $(j',b) \succ_i \mu(j,b)$ and, thus, $\mathcal{M}_i^b = j' \cup \mathcal{M}_i^b$, so the price $p_{ij}^b$ decreases since the supply increases for the existing demand.

- By contrast, the arrival of a new SU $i'$ is a *negative externality* if $U_{ij}^b$ decreases due to any of the following reasons: a) $\exists j'$, $j' \in \mathcal{M}_i^b(\mu) / (i',b) \succ_{j'} \mu(i,b)$, so this new association may reduce the quality of the connection between SU $i$ and PU $j$, or, b) $(i',b) \succ_j \mu(i,b)$ and, thus, $\mathcal{N}_j^b = i' \cup \mathcal{N}_j^b$, so the price $p_{ij}^b$ increases since the demand increases for the existing supply.

The effects are the opposite for a PU $j$. The pricing mechanism in (29) captures the previous behavior. After a price is obtained for each potential match, the SUs and PUs need to update their list of preferences since they may have changed as result of the previous externalities.

Solutions for finding a stable matching such as the deferred acceptance algorithm [32] may not converge to a stable matching when the game has externalities. Instead, we prove that our new matching algorithm converges to a stable matching to solve the data and spectrum trading game.

Let us first restate the general definition of stable matching in our game:

**Definition 3.** A local CDNA served by PU $j$ is stable if both of the following conditions are satisfied:

1) If SU $i \notin \mathcal{N}_j^b(\mu)$, then it cannot join the local CDNA. That is, there is no pair $(i, j) \notin \mu$ in which $(i,b) \succ_j (\mu(j,b),b)$ and $(j,b) \succ_i (\mu(i,b),b)$ or $(i,b) \succ_j (\mu(j,b'),b')$ and $(j,b) \succ_i (\mu(i,b'),b')$.

2) If SU $i \in \mathcal{N}_j^b(\mu)$, then it cannot leave the local CDNA. That is, there is no pair $(i, j') \notin \mu$, $j \neq j'$ in which $(j',b) \succ_i (\mu(i,b),b)$ and $(i,b) \succ_{j'} (\mu(j',b),b)$ or $(j',b) \succ_i (\mu(i,b'),b')$ and $(i,b) \succ_{j'} (\mu(j',b'),b')$.

Next, we show that the algorithm converges to a stable matching as previously defined even with externalities.

**Theorem 1.** The stability of each local CDNA is achieved after a finite number of iterations and, thus, Algorithm 3 is guaranteed to reach a stable matching and price for data and spectrum trading.

*Proof.* The proof follows from two considerations. First, SUs can reach a limited number of PUs in their vicinity due to restricted transmission ranges, and thus the number of alternatives for both SUs and PUs is finite. Indeed, each PU has a finite number of SUs to form a local CDNA and each SU has a limited number of CDNAs to switch to. Second, the price updated as in (29) converges to a stable price [33]. The stability of the price depends mainly on the learning rate. The most common way of analyzing stability is to consider the eigenvalues of the Jacobian matrix of the pricing function in (29). Following [33], the fixed point $p_{ij}^b$ is stable if and only if,

$$0 < \varpi_j < 1 + a_{ij}^b (p_{ij}^b)^2 \eta (1-\sigma) / (\rho_{ij}(\eta + 1 - \sigma)) \quad (30)$$

Recall that this is the criterion used in the selection of $\varpi_j$ in Algorithm 3.

*D. Computational Complexity and Signaling Overhead*

In the first step, each SU and PU builds its set of preferred counterparts per channel of size $|\mathcal{M}_i^b| = MB$ and $|\mathcal{N}_j^b| = NB$, respectively, by calculating the utilities (23) and (24) for each potential matching. The complexity of calculating the utilities and ordering the preferences for $N$ SUs and $M$ PUs is $\mathcal{O}(NMB(\log(MB) + \log(NB)))$. In the worst case each SU will issue $MB$ proposals to find a suitable PU. Thus, the total attempts by $N$ SUs will be at most $NMB$. Finally, the price is computed in (29), which has complexity $\mathcal{O}(NMB)$, and each PU and SU will reorder its preferences. The algorithm will terminate after a finite number of iterations $I_3$. The worst case complexity of Algorithm 3 is thus $\mathcal{O}(NMBI_3(\log(MB) + \log(NB)))$, which is significantly lower than that of Algorithms 1 and 2.

SUs and PUs exchange their respective demand and supply per channel with their selected counterparts to calculate the price, build their preference lists and avoid channel allocations that might interfere with existing matchings. As mentioned previously, in the worst case the signaling complexity is $\mathcal{O}(NMBI_3)$. The operators supervise the trading and ensure that the transactions are reflected in SUs' and PUs' monthly bills.

VI. TRUST RELATIONSHIPS IN CDNA

In a real network, SUs and PUs may act selfishly and misbehave with the aim of obtaining greater profit. In this section, we consider trust relationships between SUs and PUs to build a secure trust network for collaborative data and spectrum trading. In this regard, we develop a trust mechanism together with a behavioral-based access control scheme to incentivize and penalize honest and dishonest behavior, respectively.

Let us assume that if PU $j$ is selfish it may decide not to transmit the agreed-on data to save battery power degrading the initial utility of SU $i$, $U_{ij}^b$, to $\tilde{U}_{ij}^b$. We define the reliability of PU $j$ as the consistency of its trading agreement given by $\xi_j = \tilde{U}_{ij}^b / U_{ij}^b$. A selfish SU $i$ may also decide to leave the network before paying for the data transmitted degrading the utility of PU $j$, $V_{ji}^b$, to $\tilde{V}_{ji}^b$. Similarly, the reliability of SU $i$ is $\xi_i = \tilde{V}_{ji}^b / V_{ji}^b$. This calculation considers the most recent experience with the other party. The calculation of the reliability based also on past experiences can be obtained using the exponential moving average as in [38]. The trust of user $i$ in the connection provided by user $j$ is denoted by $\rho_{ij}$ and is included in the utility function of SU $i$, $U_{ij}^b$, in (12), (16) and (23). The trustworthiness of the connection is computed as follows,

$$\rho_{ij} = \omega O_{ij}^{dir} + (1-\omega) O_{ij}^{ind} \quad (31)$$

where $O_{ij}^{dir}$ is user $i$'s direct experience with $j$, $O_{ij}^{ind}$ is the recommendation of other users based on their experiences with $j$ and $\omega$ is the weight. All these parameters lie in the range [0,1]. Likewise, the trust of user $j$ in the connection requested by $i$ is denoted by $\rho_{ji}$ as in the utility of PU $j$, $V_{ji}^b$, in (14), (17) and (24), and is obtained as before.

We model the direct observation by user $i$ of the behavior of user $j$ using a *sigmoidal* utility function (Fig. 3) from behavioral economics [34] used to describe users' investment decisions, and extend it to model users' trading decisions in CDNA (i.e., how PUs and SUs make decisions about selling and buying resources based on the potential value of losses and gains). Recall that in CDNA connectivity is provided by users who share the residual connectivity of their devices. We assume that each user advertises its reliability level to potential counterparts. Then, direct observation by user $i$ of behavior by user $j$ is,

$$O_{ij}^{dir}(\xi_j) = 1/(1 + e^{-h(\xi_j - \phi \xi_i)}) \quad (32)$$

where $h$ and $\phi$ are scaling coefficients, $\xi_j$ is the reliability of user $j$ and $\phi \xi_i$ is the reference point. User $i$ perceives the values $O_{ij}^{dir}(\xi_j) > \phi \xi_i$ as gains and $O_{ij}^{dir}(\xi_j) < \phi \xi_i$ as loses. Note that the values of the function on both sides of the reference point are asymmetrical to capture loss aversion – people's tendency to prefer avoiding losses rather than acquiring equivalent gains.

Malicious users may lie and give their counterparts a bad recommendation (bad-mouthing) in order to have exclusive access to resources. Consequently, users will give more weight to their own observations than to others'. Besides, to increase the robustness of the trust evaluation we define the credibility of the recommender $C$ based on the number of transactions $n$ between both nodes. Thus, the credibility of the recommendation given by user $i'$, $i' \neq i$, to user $j$ is

$$C_{i'j} = n_{i'j} / (n_{i'j} + \sum_{j' \neq j} n_{i'j'}). \quad (33)$$

This expression is used to weight the indirect opinions provided by other users proportionally to the number of transactions conducted. Since SUs may have different connectivity requirements which may result in different expectations, we define a similarity factor $S_{ii'}$ between SU $i$ and $i'$ to indicate the similarity between two users and the relevance of the recommendation.

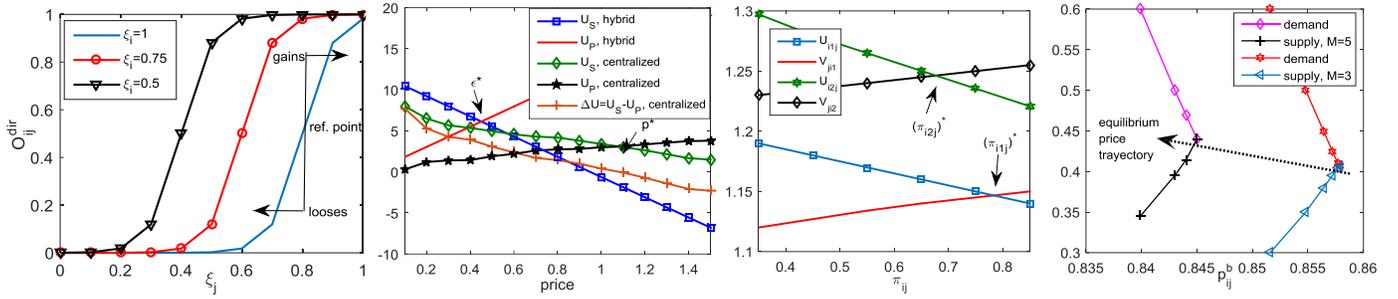

Fig.3. $O_{ij}^{dir}$ vs $\xi_j$ with $h = 20$, $\phi = 4/5$. Fig.4. Price negotiation between PO & SO. Fig.5. Price negotiation between SUs $i_1$, $i_2$ & PU $j$. Fig.6. Distributed pricing.

Thus, the recommendation is obtained as

$$O_{ij}^{ind} = \sum_{i' \neq i} S_{ii'} C_{i'j} O_{i'j}^{dir} \quad (34)$$

where $O_{i'j}^{dir}$ is observation by user $i$ of behavior by user $j$ as in (32). It is worth noting that the trustworthiness is asymmetric, i.e., $\rho_{ij} \neq \rho_{ji}$. To improve the robustness of the network to malicious users and encourage consistent behavior, we define *a behavioral-based access control mechanism* to incentivize and penalize honest and dishonest behavior by users. The PO and SO, respectively, regulate PU and SU access to CDNA by dynamically adjusting the revenue sharing parameters $\eta$ and $\sigma$ according to their users' trustworthiness. The higher the trustworthiness of the connection provided (requested) by PUs (SUs), the higher the revenue $\eta$ ($\sigma$) offered by the PO (SO). Consequently, the PO compensates participation by PU $j$ with a revenue share

$$\eta_j = \sum_i \rho_{ij} / |\rho_{ij}| \quad (35)$$

Similarly, the SO compensates SU $i$ with a revenue share

$$\sigma_i = \sum_j \rho_{ji} / |\rho_{ji}| \quad (36)$$

Both operators may also set up a trustworthiness threshold and restrict access to users who satisfy that threshold.

This two-level trust mechanism which encompasses partially distributed (via local physical interactions) and partially centralized (via the involvement of operators) trust management allows the construction of a robust and secure trust network for data and spectrum trading purposes completing the model described in Fig. 2.

## VII. Performance Evaluation

We present some numerical results to verify our theoretical analysis, evaluate the schemes and compare them with existing mechanisms. The experimental environment is Matlab.

### A. Settings

We consider a CDNA consisting of $M$ PUs and $N$ SUs randomly distributed in a 1000 x 1000 $m^2$ area. We assume that each user has a monthly contract for a data volume of 10 GB. The transmission range and interference range are set to 500 $m$ [29]. The path loss exponent is $\alpha = 4$ and $\beta = 62.5$. The noise power spectral density is $\gamma = 3.34 \times 10^{-20}$W/Hz at all nodes. The transmission power spectral density of the nodes is $8.1 \times 10^9 \gamma$, and the reception threshold and interference threshold are both $8.1\gamma$ on each spectrum band. The minimum SINR requirement varies between [5, 20] dB and the duration of the connectivity varies between [0, 10] minutes. The convergence error $\chi$ and $\chi'$ are set to $10^{-4}$.

Regarding the return of PUs, the availability $a_{ij}^b$ of a licensed band for SU transmission at a certain location has a random probability within (0.5, 1]. We run Monte Carlo simulations and average the results over 100 iterations.

### B. Data and Spectrum Trading Schemes

We assume $M = 5$ PUs and $N = 10$ SUs, a revenue share of $\sigma = \eta = 0.7$ and $B = 5$ available channels. The price negotiation process between the PO and SO is shown in Fig. 4 for the centralized and hybrid schemes. The negotiation in the distributed scheme follows the same tendency and it is omitted for clarity of presentation. The price is initialized to 0.1. In the centralized scheme, the PO and SO negotiate the price $p_{ij}^b$ per unit of data transmitted on channel $b$. For clarity of illustration, we have assumed the same price for all channels/links $p_{ij}^b = p$. The optimum trading price $p^*$ is obtained when the cooperation agreement is met, i.e., $|\Delta U| = |U_S - U_P| \leq \chi$, as shown in Fig. 4. A lower (higher) price may discourage PO (SO) participation in the trading. Similarly, in the hybrid scheme the PO and SO negotiate the price per channel $\varepsilon_{ij}^b = \varepsilon$. In Fig. 5 we present the negotiation of the data price $\pi_{ij}$ between SUs $i_1$, $i_2$ with heterogeneous requirements and PU $j$. The optimum price is obtained when $|U_{ij} - V_{ji}| \leq \chi'$ for $i = i_1, i_2$ with $\chi' \approx 0$. It is worth noting that since SUs $i_1$ and $i_2$ have different requirements the prices for their data transmission $(\pi_{i_1j})^*$ and $(\pi_{i_2j})^*$ are different. The equilibrium price in the distributed scheme is shown in Fig. 6 for $M = 3$ and 5. Each pair of demand/supply curves represents a local CDNA. As expected, increasing the price decreases the demand and increases the supply until the equilibrium price is obtained. Besides, the higher $M$ (number of supply sources) the lower the equilibrium price. In Fig. 7 we plot the overall utilities of the SUs, PUs, SO and PO for the three schemes. In the hybrid and distributed schemes, the PU and SU reach an agreement for the trading and, thus, both utilities are the same. In the centralized scheme the utility of the PU is higher than that of the SU as the PU is additionally rewarded for transmitting SU data. In the hybrid scheme the PU and SU only negotiate the price for the data.

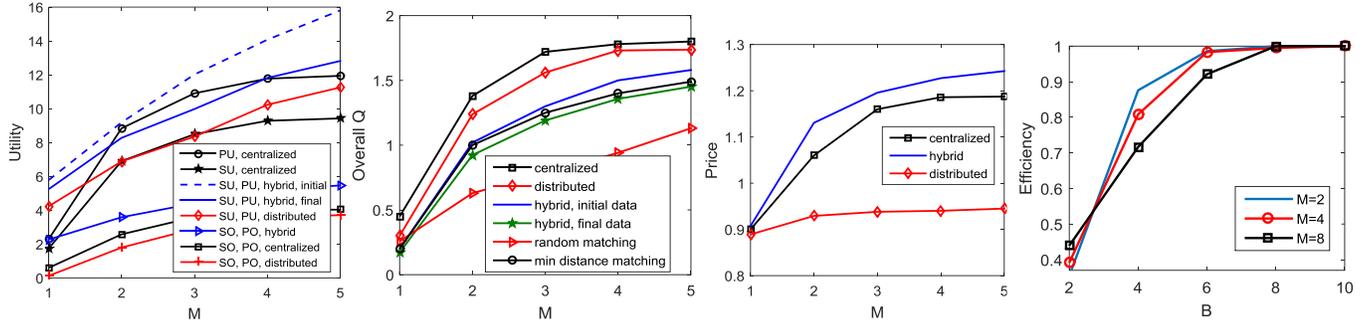

Fig. 7. $U_{SU}$, $U_{PU}$, $U_{SO}$ and $U_{PO}$ versus $M$.  Fig. 8. Overall $Q$ transmitted versus $M$.  Fig. 9. Traded price versus $M$.  Fig. 10. Trading efficiency of the hybrid scheme.

Table IV. Dynamic reconfiguration

| scenario | $N$ | $M$ | $B$ | Centralized | | Hybrid | | Distributed | | MCM | | Random | |
|---|---|---|---|---|---|---|---|---|---|---|---|---|---|
| | | | | $U_{SU}$ | CPU time (s) | $U_{SU}$ | CPU time (s) | $U_{SU}$ | CPU time (s) | $U_{SU}$ | CPU time (s) | $U_{SU}$ | CPU time (s) |
| 1 | 10 | 3 | 5 | 9.6490 | 7.6702 | 11.6007 | 0.0867 | 7.8907 | 0.0961 | 6.9810 | 0.0532 | 6.3798 | 0.00364 |
| 2 | 9 | 3 | 5 | 8.7932 | 6.0409 | 8.7597 | 0.1244 | 8.0221 | 0.0225 | 7.1601 | 0.0310 | 6.0919 | 0.00293 |
| 3 | 9 | 3 | 4 | 7.6563 | 3.7976 | 7.6430 | 0.0672 | 6.7429 | 0.0819 | 5.4280 | 0.0497 | 4.0461 | 0.00210 |
| 4 | 9 | 4 | 4 | 7.8919 | 6.6540 | 8.7443 | 0.1226 | 8.3570 | 0.0791 | 7.1674 | 0.0558 | 6.3152 | 0.00468 |
| 5 | 10 | 4 | 4 | 8.3441 | 7.4472 | 8.8493 | 0.1126 | 10.145 | 0.0725 | 9.7857 | 0.0651 | 9.1327 | 0.00423 |
| 6 | 10 | 4 | 5 | 8.9101 | 21.3824 | 11.6349 | 0.1021 | 8.5382 | 0.0386 | 8.3021 | 0.0584 | 9.7019 | 0.00483 |

Since the negotiation does not consider channel availability, the utility is higher than in the centralized scheme. After channel allocation, the final data transmitted is lower but the price remains the same, which reduces the utility. In the distributed scheme, the PU and SU negotiate the price for data and channels. With the same revenue share, their utilities are slightly higher than that of the SU utility in the centralized scheme. The utilities of the SO and PO are shown for the three schemes as well after the negotiation agreement. In the centralized scheme, both operators agree on a fixed price for the resources, while in the hybrid and distributed scheme, the PO and SO benefit from the amount of data traded proportionally to $1 - \eta$ and $1 - \sigma$, respectively. The highest utility is obtained with the hybrid scheme as this has the highest price per unit of data and spectrum.

In Fig. 8 the overall amount of data transmitted is plotted for each scheme. The highest amount of data is transmitted in the centralized scheme as the SO reuses available channels to maximize its utility. In the hybrid scheme, the PU-SU association is performed locally for data trading without information on channel availability. Allocating the channels a posteriori is less efficient and therefore less data is transmitted. In the distributed scheme, SU-PU associations are performed in a self-organized manner with local knowledge of the available resources (channels and PUs). Thus, the overall data transmitted is higher than in the hybrid scheme. Even though not all potential associations are accomplished, the overall data transmitted is slightly lower than in the centralized scheme.

These results are compared to random matching, which includes random SU-PU associations and random channel allocation. We can observe that the latter scheme is very inefficient. Besides, we also consider a minimum distance matching (MDM) scheme, which considers SU-PU associations based on shortest distance. Since it jointly produces the SU-PU associations and channel allocation it performs slightly better than the hybrid scheme but 30% and 20% worse than the centralized and distributed schemes, respectively.

In Fig. 9 the trading price per unit of data and channel is shown for the three schemes. The highest price for data and channels is obtained in the hybrid scheme. Here the price is negotiated for a higher volume of data than that finally transmitted, which reduces $Q_{aj}$ and consequently increases the price with respect to the centralized scheme, i.e., $\pi_{ij} > \Phi_{oj} / Q_{aj}$. The price for the distributed scheme is locally adjusted based on demand and supply. Since the supply of data per local CDNA is greater than the demand, the price is reduced to attract more SUs.

The trading efficiency of the hybrid scheme is shown in Fig. 10 for different numbers of channels. The efficiency is defined as the ratio between the final data traded considering channel availability and the initial data agreed-on for trading. Even with an efficiency of 1 the overall data traded in the hybrid scheme is the lowest of the three schemes. This is because the data trading association is marked by uncertain channel availability and the same association is later used for the channel allocation.

### C. Dynamic Data and Spectrum Trading

In this section, we evaluate the performance of our algorithms under traffic dynamics. The results are shown in Table IV for different scenarios, which represent different observation instants. We consider one traffic change per scenario (e.g., an SU or PU arrives or leaves or a new channel becomes available or unavailable). In the first scenario, $N = 10$, $M = 3$ and $B = 5$. In the second one, an SU leaves the network. Similar changes take place in the other cases. The computational time of the hybrid scheme is one order of magnitude faster than that of the centralized scheme but the distributed scheme reconfigures the network two orders of magnitude faster than in the centralized scheme for most scenarios considered, and even three orders of magnitude faster when there are more resources available.

The low computational time of the matching algorithm makes it a promising option for tracking traffic dynamics and reconfiguring the network in real time. The results are compared to MCM and random matching. MCM is about 15% inferior compared to the distributed scheme and its computational time is about 20% lower. Meantime, the random matching has the lowest computational time, although its performance may drop a 40% compared to that of centralized scheme.

In order to make a fair comparison of the schemes we have considered that the length of the list of preferences for each SU $i$ and PU $j$ is $\left|\mathcal{M}_i^b\right|=M$ and $\left|\mathcal{N}_j^b\right|=N$, respectively. Nevertheless, the length of the list of preferences can be limited to facilitate faster reconfigurability, and the algorithm can be stopped at any time related to desirable complexity and performance tradeoffs. Algorithm 4 incorporates the dynamic tracking of traffic variations into the matching algorithm.

**Algorithm 4** Dynamic Data and Spectrum Trading
1: Run **Stage 1 – Initialization** ($\delta = 0$)
2: Run **Stage 2 – Matching and price determination** (steps 20-22)
  SUs associate with their preferred PU and vice versa, on matching $\mu$
3: **while** $\delta < \delta_{max}$
4:   Run **Stage 2 – Matching and price determination** (steps 20-22) and obtain matching $\mu'$
5:   **if** the matching is stable ($\mu = \mu'$)
6:     reconfigure the network with $\mu'$
7:   **end**
8:   $\mu \leftarrow \mu'$; $\delta = \delta + 1$
9: **end**

*D. Trust relationships*

We evaluate the impact of trust relationships on the price and overall performance of CDNA. We use the distributed algorithm with $N = 10$ SUs, $M = 10$ PUs and $B = 10$ channels. We assume that if a user is reliable its reliability probability $\xi$ will vary randomly between [0.9, 1] and if it is unreliable it will vary between [0.5, 0.9]. SUs and PUs will be willing to connect to their counterparts when the trustworthiness of the link is $\rho_{ij}, \rho_{ji} > 0.5$, respectively. The price of data and channels is shown in Fig. 11 for different probabilities $R_i$ and $R_j$ of having a reliable SU and PU, respectively. If $R_i$ decreases, the incentive provided by the SO also decreases to penalize SU $i$. Consequently, this reduces demand and the price decreases to attract more SUs. The opposite behavior is observed for $R_j$ since a decrease in $R_j$ reduces the supply, which will increase the price.

As already mentioned, the PO and SO control PU and SU access to CDNA by dynamically adjusting parameters $\eta$ and $\sigma$ to their behavior, as shown in Fig. 12. We assume that $R_i = 0.5$. It is worth noticing that when $R_j > R_i$, the PO will encourage the PUs to join CDNA with $\eta > \sigma$ since the trustworthiness is $\rho_{ij} > \rho_{ji}$. The opposite behavior can be observed when $R_j < R_i$. Thus, the access control mechanism captures the users' behavior.

In a large CDNA, computation of trustworthiness should be autonomous to reduce the latency and overhead of exchanging recommendations. The main properties of trust (asymmetry, transitivity and composability) can be explored to automate the trust evaluation process. Regarding the property of transitivity, if $i$ trusts $j$ and $j$ trusts $i'$ then $i$ can trust $i'$. We can exploit this property and obtain $i'$ recommendation on $j$ given the similarity $S_{ii'}$ between $i$ and $i'$. Composability is the ability to compose recommendations from different users. By using the previous properties, trustworthiness is expressed as

$$\rho_{ij} = \sum_{j' \neq j / \rho_{ij}, \rho_{j'i'} > 0} \sum_{i' \neq i} S_{ii'} \rho_{i'j} / |\rho_{i'j}| \quad (35)$$

where user $i$ relies on the recommendations from every user $i'$ that has used the same PU $j'$.

In Fig. 12 we also present the accuracy of the estimation of the access control parameters when the trust calculation is automated versus $R_j$. We can see the values of $\eta$ and $\sigma$ obtained when the system is initialized with 50 and 75 samples and runs autonomously for 100 samples, compared to when it is run non-autonomously for 100 samples. In the worst case there is an error in the estimation of about 2%. It was observed that such a deviation in the access control parameters resulted in insignificant differences in terms of the utilities of users and operators.

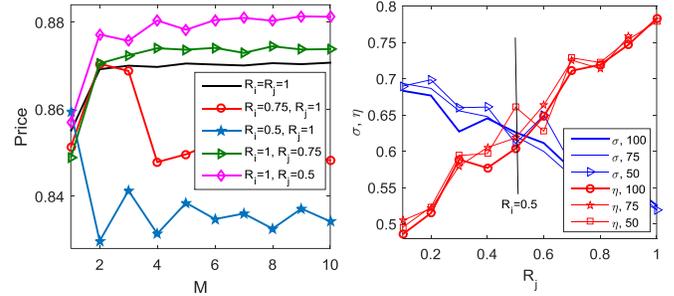

Fig. 11. Data and channel price vs. $M$.  Fig. 12. Autonomous access control.

## VIII. CONCLUSION

This paper presents collaborative trading schemes for data and spectrum sharing that create business opportunities for users and operators, while enforcing trustworthy relations. The business opportunities result from harvested data and spectrum trading between users and operators. Unlike existing approaches which mainly focus on spectrum access, by defining joint data and spectrum trading policies, we involve users in service delivery. The three schemes proposed, centralized, hybrid and distributed, progressively delegate trading to end users to favor distributed implementations. The control knobs that determine these implementations are the revenue sharing policies $\eta$ and $\sigma$ in the primary and secondary network.

Numerical results show that the hybrid scheme is 1 order of magnitude faster than the centralized scheme and that the matching algorithm reconfigures the network two orders of magnitude faster than the centralized scheme in most scenarios considered and even three orders of magnitude faster when there are more resources available. This makes the matching algorithm a promising option for exploiting available resources in real time. In addition, the performance of the hybrid scheme is very close to that of the centralized scheme. By modeling user behavior in the access control mechanism we preserve a trustworthy and autonomous trading system.

Data and spectrum trading in CDNA involves many transactions between the PO and SO and their respective users. Existing blockchain payment systems can facilitate secure transactions although reducing consensus latency in very dense networks is an open challenge. Besides, blockchain payments can provide reliable information on recorded transactions contributing to trustworthiness in CDNA. This work could be used as a case study to develop simplified payment systems based on blockchain technology. Besides, it opens future research opportunities such as developing mathematical frameworks for reasoning about trust, modeling of user misbehavior, and its extension to real traffic models.

## APPENDIX

*Proof of convergence for Algorithm 1:*

In the following we prove that after successive negotiations the PO and SO reach an agreement i.e., $|\Delta U| = |U_S - U_P| \leq \chi$ and, thus, Algorithm 1 converges to the optimum price $(p_{ij}^b)^*$ for each SU $i \in \mathcal{N}$ and PU $j \in \mathcal{M}$ in channel $b \in \mathcal{B}$. For simplicity we assume that users transmit the same amount of data $Q_{ij} = Q$ and all links have the same price $p_{ij}^b = p$, and $n = \sum_i \sum_j \sum_b t_{ij}^b$. Let us rewrite the price update in a compact form as: $p(t+1) = p(t) + \Delta p \cdot \text{sgn}(\Delta U(t))$ with $\text{sgn}(\Delta U(t)) = -1$ if $\Delta U(t) < 0$, $\text{sgn}(\Delta U(t)) = 0$ if $\Delta U(t) = 0$, and $\text{sgn}(\Delta U(t)) = 1$ if $\Delta U(t) > 0$. If $\Delta U(t) > 0$, the new price $p(t+1)$ increases moving in the direction towards the agreement point. On the other hand, if $\Delta U(t) < 0$, the new price $p(t+1)$ decreases until the agreement point is reached. Thus, the function $\Delta U$ is monotone decreasing i.e. $\partial \Delta U(p)/\partial p < 0 \ \forall p$ and the system converges eventually to a single global equilibrium point. The step size required to bring the positions closer i.e. $|\Delta U(t+1)| < |\Delta U(t)|$ in each case is: a) if $\Delta U(t) > 0$ after the price update $U_S(t+1) < U_S(t)$. The new PO utility must satisfy $U_P(t+1) \geq U_P(t)$ and thus the required price step is $\Delta p \geq (p - \eta \Phi_o / Q_a)(nQ/n'Q'-1)$; b) if $\Delta U(t) < 0$ after the price update explained above $U_S(t+1) > U_S(t)$. To reduce the agreement gap, the new PO utility must satisfy $U_P(t+1) \leq U_P(t)$ and therefore the price step $\Delta p \geq (p - \eta \Phi_o / Q_a)(1 - nQ/n'Q')$. After successive iterations the agreement price is reached and the optimum price is $p^* = (\rho \log(Q_o + Q)/Q + \eta \Phi_o / Q_a)/2$.

## ACKNOWLEDGEMENT

This work was funded by Fulbright Program CAS16/00093 and grants TEC2016-76465-C2-2-R, IJCI-2014-20611 and EUIN2017-88225, MINECO, Spain. The work of A. S. Shafigh was supported by the Finnish Academy/NSF US collaborative program/WiFiUS 2018, and the work of Y. Fang was partially supported by the National Science Foundation under grants CNS-1717736 and CNS-1343356.